\documentclass[submission,copyright,creativecommons]{eptcs}
\providecommand{\event}{F-IDE 2015} 
\usepackage{breakurl}             

\usepackage{amssymb}
\usepackage{graphicx}

\usepackage{url}
\usepackage{cite}

\urldef{\mailsa}\path|{nokovib, emil}@mcmaster.ca|

\usepackage{url}
\usepackage{mathptmx}
\usepackage{amsmath}
\usepackage{amssymb}
\usepackage{amsfonts}
\usepackage{latexsym}
\usepackage{hyperref}
\usepackage{tikz}
\usepackage{graphicx}
\usepgflibrary{shapes.geometric}
\usepgflibrary{shapes.misc}
\usetikzlibrary{topaths} 			 
\usetikzlibrary{automata} 			 
\usetikzlibrary{backgrounds} 		
\usepgflibrary{decorations.markings}
\usetikzlibrary{shadows}
\usepgflibrary{shapes.symbols} 	
\usetikzlibrary{er}
\usetikzlibrary{chains}
\usepackage[scaled=.92]{helvet}
\usepackage{graphicx}
\usepackage{listings}
\lstnewenvironment{code}{}{}
\usepackage{float}
\usepackage{comment}
\usetikzlibrary{positioning,arrows}

\def\gpicbox#1{\vbox{\unvbox\csname #1\endcsname\kern 0pt}}

\usepackage[nounderscore]{syntax}
	
\usepackage{url,mathptmx,listings,hyperref,amsmath}
\usepackage{latexsym}
\usepackage[scaled=.92]{helvet}

\catcode`\@=11
\SetMathAlphabet{\mathrm}{normal}{\encodingdefault}{\rmdefault}{m}{n}%
\SetMathAlphabet{\mathbf}{normal}{\encodingdefault}{\rmdefault}{bx}{n}%
\SetMathAlphabet{\mathsf}{normal}{\encodingdefault}{\sfdefault}{m}{n}%
\DeclareSymbolFont{italics}{\encodingdefault}{\rmdefault}{m}{it}%
\DeclareSymbolFontAlphabet{\mathrm}{operators}
\DeclareSymbolFontAlphabet{\mathit}{letters}
\DeclareSymbolFontAlphabet{\mathcal}{symbols}
\def\@setmcodes#1#2#3{{\count0=#1 \count1=#3
        \loop \global\mathcode\count0=\count1 \ifnum \count0<#2
        \advance\count0 by1 \advance\count1 by1 \repeat}}
\@setmcodes{`A}{`Z}{"7\hexnumber@\symitalics41}%
\@setmcodes{`a}{`z}{"7\hexnumber@\symitalics61}%
\newcommand\normalfootnotesize{%
   \@setfontsize\normalfootnotesize\@viiipt{9.5}%
   \abovedisplayskip 6\p@ \@plus2\p@ \@minus4\p@
   \abovedisplayshortskip \z@ \@plus\p@
   \belowdisplayshortskip 3\p@ \@plus\p@ \@minus2\p@
   \def\@listi{\leftmargin\leftmargini
               \topsep 3\p@ \@plus\p@ \@minus\p@
               \parsep 2\p@ \@plus\p@ \@minus\p@
               \itemsep \parsep}%
   \belowdisplayskip \abovedisplayskip
}

\font\msx=msam10
\font\msy=msbm10

\newfam\msxfam \textfont\msxfam=\msx
\newfam\msyfam \textfont\msyfam=\msy

\def\famletter#1{\ifcase #1 0\or 1\or 2\or 3\or 4\or 5\or 6\or 7\or
	8\or 9\or A\or B\or C\or D\or E\or F\fi}

\edef\fx{\famletter\msxfam}

\def\@myop#1{\mathop{\mathstrut{#1}}\nolimits}

\def\_{\leavevmode \vbox{\hrule width0.5em}}

\let\xforall=\forall
\let\xexists=\exists
\let\xlambda=\lambda

\let\mc=\mathchardef

\def	\po1		{\mbox{${\cal P}_1$}}

\def	\lambda		{\@myop{\xlambda}}

\def	\forall		{\@myop{\xforall}}
\def	\exists		{\@myop{\xexists}}

\def	\dom		{\@myop{\sf dom}}
\def	\ran		{\@myop{\sf ran}}
\def	\id		{\@myop{\sf id}}
\def	\comp		{\mathbin{\raise
			0.6ex\hbox{\oalign{\hfil$\scriptscriptstyle
			\rm o$\hfil\cr\hfil$\scriptscriptstyle\rm 9$\hfil}}}}

\mc	\dres		"2\fx43
\mc	\rres		"2\fx42

\mc	\inj		"3\fx1A

\mc	\surj		"3\fx10

\def	\na1		{\mbox{${\cal N}_1$}}

\def	\int		{\mbox{${\cal Z}$}}

\def	\finse1		{\mbox{${\cal F}_1$}}

\def	\seq		{\@myop{\rm seq}}
\def	\cat		{\mathbin{\raise 0.8ex\hbox{$\mathchar"2\fx61$}}}

\catcode`\@=12

\newcommand{\choice}{\mathrel{[\hspace{-0.5pt}]}}

\newcommand{\IN}{\mathop{\sf in}}

\usetikzlibrary{shapes.callouts}

\pgfdeclarelayer{background}
\pgfdeclarelayer{foreground}
\pgfsetlayers{background,main,foreground}
\tikzstyle{reqm}=[draw, text width=6.15em, 
    text centered, minimum height=3.5em]
\tikzstyle{reqm1}=[draw, text width=9.15em, 
    text centered, minimum height=3.5em]
    
\tikzstyle{ann} = [above, text width=5em, text centered]
\tikzstyle{wa} = [reqm, text width=8em, fill=white!20, 
    minimum height=3.5em, rounded corners]  
\tikzstyle{em} = [reqm, text width=8em, fill=white!20, 
    minimum height=6.9em, rounded corners]    
\tikzstyle{mc} = [reqm1, text width=8em, fill=white!20, 
    minimum height=3.9em, rounded corners]       
\tikzstyle{sc} = [reqm, text width=13em, fill=red!20, 
    minimum height=10em, rounded corners, drop shadow]  
\tikzstyle{gcode} = [draw, fill=white, text width=4em, 
    text centered, minimum height=3.5em]

\title{A Holistic Approach in Embedded System Development}

\author{Bojan Nokovic
\institute{Computing and Software Department\\
McMaster University\\
Hamilton, Canada}
\email{nokovib@mcmaster.ca}
\and
Emil Sekerinski
\institute{Computing and Software Department\\
McMaster University\\
Hamilton, Canada}
\email{\quad emil@mcmaster.ca}
}

\def\titlerunning{A Holistic Approach in Embedded System Development}

\def\authorrunning{B. Nokovic \& E. Sekerinski}

\begin{document}
\maketitle

\begin{abstract}
We present \emph{pState}, a tool for developing ``complex'' embedded systems by integrating validation into the design process. The goal is to reduce validation time. To this end, qualitative and quantitative properties are specified in system models expressed as pCharts, an extended version of hierarchical state machines. These properties are specified in an intuitive way such that they can be written by engineers who are domain experts, without needing to be familiar with temporal logic. From the system model, executable code that preserves the verified properties is generated. The design is documented on the model and the documentation is passed as comments into the generated code. On the series of examples we illustrate how models and properties are specified using \emph{pState}.
\end{abstract}

%
\section{Introduction}
\label{sec:Introduction}

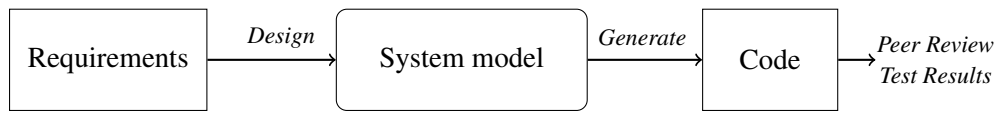
\begin{figure}[h]
\centering 
\begin{tikzpicture}
\draw (-1.5,5.1)  node[reqm](r1) {Requirements};
\draw (3.2,5.1)  node[wa](sm1) {};
\draw (3.2, 5.1) node {System model};
\draw (0.8, 5.4) node {\footnotesize{$Design$}};
\draw[thick,->] (r1) -- (sm1);
\draw (7.3,5.1)  node[gcode](c1) {};
\draw (5.6, 5.4) node {\footnotesize{$Generate$}};
\draw[thick,->] (sm1) -- (c1);
\draw (7.3, 5.1) node {Code};
\draw (9.5, 5.3) node {\footnotesize{$Peer \; Review$}};
\draw (9.5, 4.9) node {\footnotesize{$Test \; Results$}};
\draw[thick,->] (c1) -- (8.7, 5.1);
\end{tikzpicture}
\caption{Conventional Software Design Process}
\label{fig:Conventional}
\end{figure}

The main traditional software validation techniques are \emph{peer review}, \emph{testing}, and \emph{performance measurement}. Peer review is the process of static code examination by the author and colleagues. The goal is to detect and identify problems and to confirm main design decisions. Quantitative studies indicate that peer review is an effective technique which catches on average about 60\% of defects~\cite{BaierKatoen:2008}. Software testing and performance measurement examine code by executing it on a specified target in a particular environment. For each specified input, a test is performed. Correctness is determined based on program execution paths. It is often not possible to test all execution paths, especially in  concurrent systems, so in practice correctness is determined on a subset of all possible executable paths. This implies that the correctness is relative to the examined paths. The correctness of  software systems in the conventional software development process, shown in Figure~\ref{fig:Conventional}, is relative to the specification and to consequently executed test cases. This process can discover errors but cannot guarantee correctness: an error may still exist in the product. The other problem is that errors are discovered \emph{late}, when the product is already built. The sooner errors are found, the lower the cost of repairing them is.

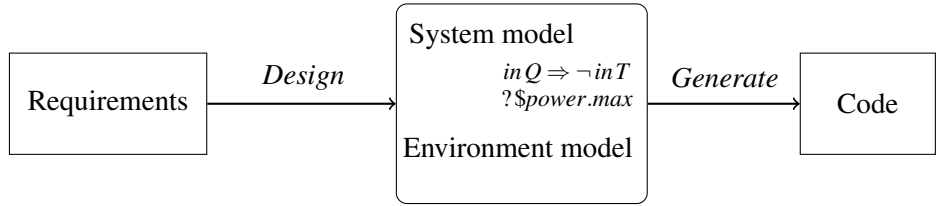
\begin{figure}[H]
\centering 
\begin{tikzpicture}
\draw (-1.5,1.1)  node[reqm](r2) {Requirements};
\draw (4.0,1.1)  node[em](sm2) {};
\draw (3.6, 2.00) node {System model};
\draw (3.95, 0.52) node {Environment model};
\draw[thick,->] (r2) -- (sm2);
\draw (8.6,1.1)  node[gcode](c2) {};
\draw[thick,->] (sm2) -- (c2);
\draw (4.6, 1.5) node (f1) {\footnotesize{$in \, Q \Rightarrow  \neg \, in \, T$}};
\draw (4.6, 1.15) node (f2) {\footnotesize{$ ? \, \$power.max $}};
\draw (8.6, 1.1) node {Code};
\draw (1.1, 1.45) node {$Design$};
\draw (6.7, 1.45) node {$Generate$};
\end{tikzpicture}
\caption{Model Checking Software Design Process}
\label{fig:ModelChecking}
\end{figure}

To overcome these two problems in the development of \emph{embedded systems}, a technique based on \emph{model checking} is proposed. From a model that describes the system behaviour in a mathematically precise manner, a simplified model suitable for model checking is generated. Model checking allows to explore all possible system states in a systematic manner. In our approach, the system model is described by pCharts~\cite{NokovicSekerinski14pCharts}, a version of hierarchical state machines extended with \emph{probabilistic transitions}, \emph{timed transitions}, \emph{stochastic timing}, and \emph{costs/rewards}. Using pCharts we can specify both a system and its environment. Qualitative and quantitative properties are expressed directly in the model. Verification of qualitative properties returns \emph{true} or \emph{false}. Verification of quantitative property returns a numerical value.  In Figure~\ref{fig:ModelChecking}, the qualitative property $"in \, Q \Rightarrow  \neg \, in \, T"$ states that whenever the system is in the state $Q$ it should not be in the state $T$. The quantitative query~$"? \, \$power.max"$ returns the maximum value of \emph{power} spent by the time a state is reached.  

In embedded systems, the impact of the working environment and the reaction to external stimuli determine the correctness of a system. In order to verify those properties, a model of the environment has to be created together with a model of the system. If errors are detected by model checking, either the system model or the environment model need to be modified. Once the properties are verified from the system model executable code that preserves those properties is generated. The same formalism, pCharts is used to describe the system model and the environment (unlike in classical discrete event systems). Code is generated according to the algorithm presented in~\cite{NokovicSekerinski14pCharts}, and tested on a number of case studies. Although it is not formally proved to be correct, we believe that generated code is \emph{trustworthy}.  

The main strength of the model checking approach is the automatic verification based on sound mathematical foundations. The main weakness is related to scalability, a.k.a. the state-space-explosion problem, and the fact that it verifies a system model, not the system itself. Because of that we can say that any model checking technique is only as good as the model of the system itself. Model checking technique enjoys a rapidly increasing interest by industry~\cite{BaierKatoen:2008} because of its potential to make the engineering process more efficient.

%
\section{Related Work}
\label{sec:RelWork}
Probabilistic transitions can be used to express randomized algorithms or quantify the uncertainty of the environment. Probabilistic descriptions are useful for analyzing quality of service, response time, unreliable environments, and fault-tolerant systems. Quantitative queries can also be attached to any state in a state hierarchy. In general, quantitative queries are specified in the temporal logic PCTL~\cite{BaierKatoen:2008} with operators for probabilities and costs/rewards. 
Analysis  in \emph{pState} proceeds by generating from the visual specification a Markov Decision Process (MDP) model and quantitative queries that are passed to a probabilistic model checker. An MDP model of PRISM~\cite{KwiatkowskaNormanParker11PRISM4} corresponds to a probabilistic timed automaton of~\cite{Segala95RandomizedDistributedRealTime}.

Modelling tools like RHAPSODY, \emph{Stateflow} with Simulink, \emph{Papyrus}, \emph{Yakindu}, SCADE Suite, and  \emph{IAR visualSTATE} do not support invariants, probabilistic transitions, stochastic transitions and direct cost/rewards specification. 
In~\cite{ZhaoYangXieLiu00QuantitativeUMLStateDiagrams} translation of extended UML diagrams{~\cite{Jansen03StatechartExtensions} to PRISM is proposed, but  parallel composed system is passed as a system of multiple models to PRISM. This process does not allow nested concurrency representation, only top level concurrency. On the other hand \emph{pState} translates arbitrarily nested parallel compositions and creates one model. Advantage is that in addition to the model checker input code, it is possible to generate executable code, so \emph{pState} is not only a PRISM front-end.    

We use PRISM for two reasons: (1) pChart's normal form of transitions correspond to the form of the input commands to PRISM, and (2) PRISM shows better overall performance compared to other probabilistic model checking tools like ETMCC, MRMC, YMER, VESTA~\cite{Oldenkamp07}. PRISM uses efficient algorithms and data structures based on binary decision diagrams that allow compact representation and increase tool scalability~\cite{KwiatkowskaNormanParker11PRISM4}.

Model checking tools based on exhaustive checking are  feasible only for systems with up to $10^{8}$-$10^{9}$ states. To address this problem, statistical model checking (SMC) which solves the verification problem for stochastic systems in a less precise but still rigorous and efficient way is introduced~\cite{Clarke:StatisticalMC}. UPPAAL tool with SMC extension is capable to verify systems with more than $10^{9}$ states, so it can verify bigger systems than \emph{pState} but in less precise manner. 

The original motivation for this work came from the design of RFID tags for postal systems~\cite{NokovicSekerinski10Interrrogator}. The use of PRISM for wireless network protocols is further studied in~\cite{Fruth11}, 
where the example in Figure~\ref{fig:SenderReceiver} is specified by a simple state diagram. Overall our thrust is to have a tool founded in solid theory and is intuitive enough to be used by engineers for analyzing design tradeoffs. An overview of~\emph{pState} with examples of generated code is in~\cite{NokovicSekerinski13pState}. In this work we focus on the property specification and generation of the documentation.\footnote{\emph{pState} can be downloaded at \url{http://pstate.mcmaster.ca}}

%
\paragraph{Requirements-oriented Semantics.}
Hierarchical state machines, pCharts, are meant primarily for the design of embedded synchronous reactive systems. Transitions in the top level states have priority over transitions in lower level states. Reset always bring the system to its initial state, and there is no history transitions. We can think of a pCharts as a compact representation of a flat-state machine. On the response to an external event, a pCharts may broadcast additional events. The execution step completes as soon as the chain of reactions comes to a halt~\cite{luttgen:00}. With embedded systems in mind, pCharts follow an \emph{event-centric} interpretation, in which events are executable procedures, implying that their execution is fast enough and  no queuing of events is needed~\cite{SZ01}. That is, if an event leads to broadcasting of another event, the second one is executed like a called procedure, rather than queued. This is in contrast to the \emph{state-centric} interpretation in UML and Statemate~\cite{HN96}, in which events are data in queues. These interpretations are called \emph{requirements-oriented} and \emph{implementation-oriented} semantics in~\cite{EshuisJansenWieringa02RequirementsStatecharts}, with our event-centric interpretation being the requirements-oriented semantics.

A number of quantitative extensions of statecharts, similar to pCharts, have been proposed, all based on UML state machines~\cite{JansenHermannsKatoen02ProbabilisticUMLStatecharts,Jansen03StatechartExtensions,LeitnerFischerLeue11QuantUM,ZhaoYangXieLiu00QuantitativeUMLStateDiagrams}. These  follow a state-centric interpretation where the state of a chart is given by its configuration (the ``states'' of the current state), a set of events, and the valuation of the variables (e.g. p. 67 in~\cite{Jansen03StatechartExtensions}). In pCharts, the state consists only of the configuration and the valuation of the variables, thus reducing the state space and facilitating model checking. UML state machines events have a single receiving object, whereas pCharts follow the original Statemate interpretation and always broadcast events to all concurrent states.

The event-centric interpretation allows pCharts to be represented through \emph{probabilistic guarded commands}~\cite{MorganMcIverSeidel96ProbabilisticPredicateTransformers}. The translation is simple and intuitive enough to serve as the definition of pCharts. The definition supports state hierarchies with inter-level transitions and concurrent states with broadcasting in arbitrary combinations. The event-centric translation of statecharts without probabilities in~\cite{Sekerinski09StateInvariants}, generates \emph{nested} guarded commands as supported in the B Method~\cite{Abrial96B,SZ01}. However, for the purpose of probabilistic model checking, a \emph{flat} structure of guarded commands is needed. A probabilistic guarded command is in \emph{normal form} if it is a nondeterministic choice among a set of guarded statements,
  \[b_1 \to S_1 \choice \cdots \choice b_m \to S_m\]
where each $b_i$ is a Boolean expression, each $S_i$ is a probabilistic choice among multiple assignment statements $A_j$ with probability $p_j$:
  \[p_1: A_1 \oplus \cdots \oplus p_m: A_m\]
Thus \emph{pState} translates pCharts to probabilistic guarded commands in normal form. A variation is used to translate sub-charts without probabilistic choice to nested control structures, which can be executed more efficiently than flat guarded commands. From an intermediate representation of nested control structures C code generation is currently supported. We extend the hierarchical chart structure to allow the specification of a cost/reward of being in a state and of taking a transition. We are not aware that costs in this form have been considered for hierarchical charts. In \emph{pState} the cost/reward specifications are first validated and then translated as annotations of the generated probabilistic guarded commands. A theory for costs/rewards in that form is given by \emph{priced probabilistic automata}; a recent overview with model checking procedures is given in~\cite{NormanParkerSproston12ModelCheckingPTA}. 

%
\section{Property Specification}
\label{sec:Ver}
The example of the wireless sender and mobile receiver in Figure~\ref{fig:SenderReceiver} illustrates the basic elements of pCharts. The setup is typical for networks of sensors, in particular RFID tags. The state \emph{System} is an AND (concurrent) state with children \emph{Sender} and \emph{Receiver}, separated by a dashed line. Both \emph{Sender} and \emph{Receiver} are XOR states with Basic states as children. The sender is initially in the state \emph{Sleep} and the receiver in \emph{Listening}. The sender exits sleep mode on a wake-up event. For active RFID tags~\footnote{\url{http://www.lyngsoesystems.com/Canada/Tags.asp}}, that event can be created either by a low frequency electromagnetic field, by a motion sensor, or by an internal timer. When this event is generated by a motion sensor or an internal timer, the sender always goes into transmission mode. On the other hand, an electromagnetic field can be created by system antennas (good field), or by other sources like power lines, monitors, cell phones, or electrical machines (parasite field). A good field has a unique identification number. If the sender recognizes the field number, it goes into transmission mode; otherwise it goes back into sleep mode. This is expressed by a probabilistic transition that with probability $0.4$ goes to state \emph{Sending} and with probability $0.6$ goes back to state \emph{Sleeping}. A sent message may reach the receiver or may get lost. This is expressed by another probabilistic transition that with probability $0.9$ broadcasts $msg$ to the receiver, which causes the receiver to go from \emph{Listening} to \emph{Off}. The receiver then shuts off to save power, while the sender (with unidirectional transmission) keeps retransmitting the message.
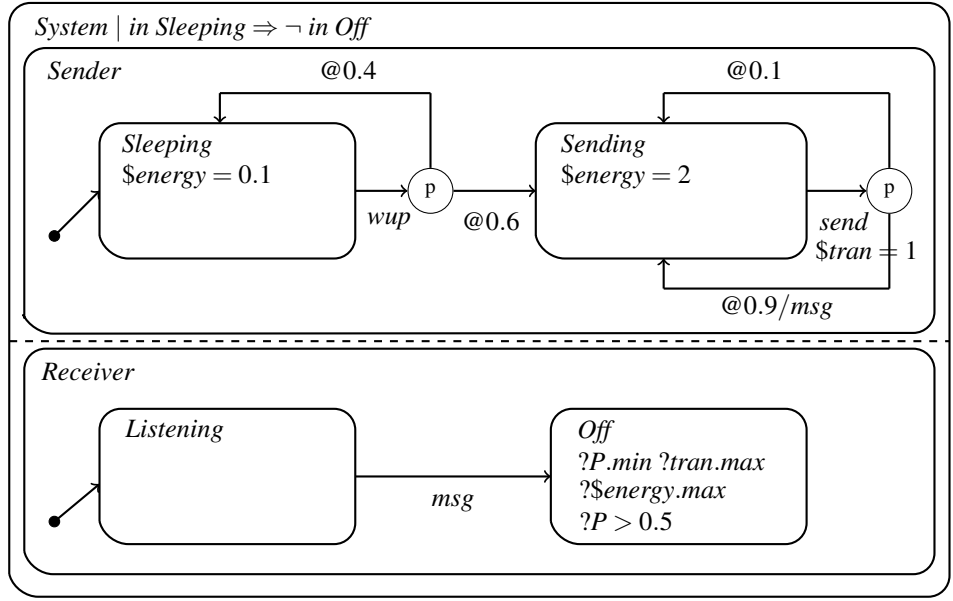
\begin{figure}[h]
\centering
\begin{tikzpicture}
        	\draw [thick,rounded corners=8pt, fill=white!16] (0.2,5.4) -- (0.2,9) -- (12.3,9) -- (12.3,5.21)--(0.3,5.2)--(0.2,5.42);
   	\draw[thick,rounded corners=8pt, fill=white!16] (0.2,2.3) -- (0.2,5) -- (12.3,5) -- (12.3,2.0)--(0.3,2.0)--(0.2,2.3);
       	\draw (2.58 ,9.26) node {\small{{$System$ $|$ $in \; Sleeping  \Rightarrow \neg \; in \; Off$}}};
 	\draw[thick,rounded corners=8pt] (0,2.0) -- (0,9.6) -- (12.5,9.6) -- (12.5,1.7)--(0.1,1.7)--(0,2.0);
	\draw [dashed, thick] (0,5.1) -- (12.5,5.1);
	\draw (1.0 ,8.7) node {\small{$Sender$}};
	\draw (1.05 ,4.7) node {\small{$Receiver$}};
	 \draw[thick,rounded corners=8pt, fill=white!30] (7.0,6.5) -- (7.0,8) -- (10.6,8) -- (10.6,6.2)--(7.1,6.2)--(7.0,6.5);
	\draw (7.9 ,7.7) node {\small{$Sending$}};
	\draw (8.2 ,7.3) node {\small{$\$energy=2$}};
	\draw (11.1 ,6.7) node {\small{$send$}};
	\draw (11.4 ,6.3) node {\small{$\$tran=1$}};
	\draw (10.2 ,5.55) node {\small{$@0.9/msg$}};
	\draw (9.9 ,8.7) node {\small{$@0.1$}}; 
	\draw[thick,rounded corners=8pt, fill=white!30] (1.2,6.5) -- (1.2,8) -- (4.6,8) -- (4.6,6.2)--(1.3,6.2)--(1.2,6.5);
	\draw (2.1 ,7.7) node {\small{$Sleeping$}};
	\draw (2.5 ,7.3) node {\small{$\$energy=0.1$}};
	\filldraw (0.6,6.5)   circle (2pt);
	\draw[thick,->] (0.6,6.5) -- (1.2,7.1) ;
	\draw (5.05 ,6.7) node {\small{$wup$}};
	\draw (6.4 ,6.7) node {\small{$@0.6$}}; 
	\draw (4.5 ,8.7) node {\small{$@0.4$}}; 	
	\draw[thick,rounded corners=8pt, fill=white!30] (7.2,2.7) -- (7.2,4.2) -- (10.6,4.2) -- (10.6,2.4)--(7.3,2.4)--(7.2,2.7);
	 \draw[thick,rounded corners=8pt, fill=white!30] (1.2,2.7) -- (1.2,4.2) -- (4.6,4.2) -- (4.6,2.4)--(1.3,2.4)--(1.2,2.7);
	\filldraw (0.6,2.7)   circle (2pt);
	\draw[thick,->] (0.6,2.7) -- (1.2,3.2) ; 
	\draw (5.9 ,2.95) node {\small{$msg$}};
	\draw (2.2 ,3.9) node {\small{$Listening$}}; 
	\draw (7.8 ,3.9) node {\small{$Off$}};
	\draw (8.8 ,3.5) node {\small{$? P.min \; ? tran.max$}};  
	\draw (8.55 ,3.1) node {\small{$? \$ energy.max$}};
	\draw (8.2 ,2.7) node {\small{$? P>0.5$}};
	\draw (5.6,7.1)  node(p1) [shape=circle,draw ] {\scriptsize{p}};
	\draw[thick,->] (4.6, 7.1) -- (p1); 
	\draw[thick,-] (p1) -- (5.6, 8.4); 
	\draw[thick,-] (2.8, 8.4) -- (5.6, 8.4); 
	\draw[thick,->] (2.8, 8.4) -- (2.8, 8);
	\draw[thick,->] (p1) -- (7.0, 7.1);
	\draw (11.7,7.1)  node(p2) [shape=circle,draw ] {\scriptsize{p}};
	\draw[thick,->] (10.6, 7.1) -- (p2); 
	\draw[thick,-] (p2) -- (11.7, 8.4); 
	\draw[thick,-] (8.7, 8.4) -- (11.7, 8.4); 
	\draw[thick,->] (8.7, 8.4) -- (8.7, 8);
	
	\draw[thick,-] (p2) -- (11.7, 5.8); 
	\draw[thick,-] (11.7, 5.8) -- (8.7, 5.8); 
	\draw[thick,->] (8.7, 5.8) -- (8.7, 6.2);
	
	\draw[thick,->] (4.6, 3.3) -- (7.2, 3.3);
\end{tikzpicture}
\caption{Sender-receiver}
\label{fig:SenderReceiver}
\end{figure}

We wish to analyze the following properties of the system:
\begin{itemize}
\item Is the system correct in the sense that the receiver is attentive when needed? We express this by attaching the invariant $(Sender \IN Sleeping) \Rightarrow \neg (Receiver \IN Off)$ to the state $System$; \emph{pState} reports $true$.
\item What is the minimal probability that the receiver shuts off? We express this by attaching the query $? P.min$ to state $Off$; \emph{pState} reports $1.0$.
\item What is the maximal number of expected message transitions of the sender until the receiver shuts off? For this, we attach the cost of $\$tran = 1$ to the sending transition and ask what the maximal expected value of $tran$ upon entering state $Off$ is by attaching the query $?\$tran.max$ to $Off$; \emph{pState} reports $1.11$.
\item What is the maximal expected energy consumption until the message reaches the receiver? For this, we attach the cost of $\$energy = 0.1$ to state $Sleep$ and $\$energy = 2$ to $Sending$. Now we can ask what the maximal expected value of $energy$ is in state $Off$ by attaching the query $?\$energy.max$ to $Off$; \emph{pState} reports $2.39$.
\item Is the probability that the receiver shuts off at least $0.5$? We express this by attaching the query $P>0.5$ to $Off$; \emph{pState} reports $true$.
\end{itemize}

In general, state invariants are \emph{safety conditions} that can be attached to any state in a state hierarchy and specify what has to hold in that state. Every incoming transition to the state must ensure that the state invariant holds, and every outgoing transition can assume that the invariant holds. State invariants can express safety of an embedded system or consistency of a software system. The \emph{accumulated invariant} of a state consists of a conjunction of invariants ``inherited'' from ancestor states and a combination of invariants of descendant states. For example, the accumulated invariant of \emph{System} is the invariant attached to \emph{System} and the invariant of \emph{Sleeping} or \emph{Sending} and the invariant of \emph{Listening} or \emph{Off}. We have implemented accumulated invariants in pCharts  following the definition and algorithm of~\cite{Sekerinski09StateInvariants}. When using a model checker for invariant verification, we interpret invariants as temporal always-conditions rather than as inductive invariants as in~\cite{Seker08}. For this example \emph{pState} generates the code in Figure~\ref{fig:Sender-receiver Code}.

\begin{figure}[!htb]
\medskip
\begin{lstlisting}[frame=single, mathescape]
mdp 

const Sending=0; const Sleeping=1; 
const Off=0; const Listening=1; 

module SenderReceiver
    sender:[0..1] init Sleeping;    
    receiver:[0..1] init Listening;    

    [wup] (sender=Sleeping) -> 0.6:(sender'=Sending) + 0.4:(sender'=Sleeping);
    [send] (sender=Sending) &(receiver!=Listening) -> 0.1:(sender'=Sending) + 
    		0.9:(sender'=Sending);
    [send] (sender=Sending) &(receiver=Listening) -> 0.1:(sender'=Sending) + 
    		0.9:(sender'=Sending)& (receiver'=Off);
endmodule

// State rewards
rewards "energy"
    (sender=Sending): 2;
    (sender=Sleeping): 0.1;
endrewards

// Transition rewards
rewards "tran"
    [send] true:1;
endrewards
\end{lstlisting}
\caption{Generated Code for Sender-receiver Code}
\label{fig:Sender-receiver Code}
\end{figure}

Queries in PRISM are specified separately from the PRISM model. In this section, \emph{state} refers to states of PRISM models, i.e.~configurations of pCharts and \emph{path} is a sequence of PRISM states. The Boolean-valued query  $P \sim r[pathprop]$, where $\sim$ is $<, <=, >, >=, =$, is true in a state if the probability that $pathprop$ is satisfied by the paths from that state is $\sim r$.  Among the path properties that can be specified is the \emph{always} property, written $G~prop$. The invariant $(Sender \, \IN \, Sleeping) \Rightarrow \neg (Receiver \IN Off)$ of state $System$ is translated by~\emph{pState} as:
   \[P>=1[G ((sender = Sleeping) => !(receiver = Off))]\]

\noindent The real-valued queries $Pmin=? [pathprop]$ and $Pmax=? [pathprop]$ return the minimal and maximal probability, which may differ due to nondeterminism. The query for the minimal probability that the receiver eventually shuts off uses the \emph{eventually} operator $F~prop$:
  \[Pmin=? [F (receiver = Off)]\]
The total reward for a path is the sum of the state rewards plus transition rewards  along the path. The Boolean-valued query $R\sim r[rewardprop]$,  evaluates to true in a state if the \emph{expected reward} associated with $rewardprop$ is ${\sim r}$ when starting from that state. The real-valued queries ${Rmin=? [rewardprop]}$ and $Rmax=? [rewardprop]$ return the minimal and maximal reward, which may again differ due to nondeterminism. For $rewardprop$ we consider only the \emph{reachability} reward $F~prop$, which is the reward accumulated along a path until a state satisfying $prop$ is reached. The maximum expected number of transmission attempts of the sender until a message reaches the receiver and the receiver shuts off is expressed as follows; as several \emph{reward structures} can be specified, reward formulae have to refer to the structure, here $tran$:
  \[R\{"tran"\}max=? [F (receiver = Off)]\]
For the maximal energy \emph{pState} generates:
  \[R\{"energy"\}max=? [F (receiver=Off)]\]

\noindent The maximal expected energy is calculated by PRISM as $2.39$. If the transitions were reliable, the result would be $2.1$.
Such analysis can be used to evaluate tradeoffs. For example, if we assume that spending 10\% more energy for sending increases the probability of successful transmission to $0.98$, we obtain that 1.02 transmission attempts are needed. This gives an appealing alternative to the practice of basing such evaluations exclusively on lab experiments.

\subsection{The Logic PCTL in pCharts} 
pCharts allows both model design and property specification in the same hierarchical state structure, while the specification of properties in the PRISM model is done separately from the model itself. To express a query decsribing minimal probability to reach some state, we need to write  formula
\[
Pmin =?[F(scopeVariable = state)]
\] 
in PRISM. In pCharts, we do not need to write full formula. By writing only~
\emph{"?P.min"} and attaching it to the state of interest we can specify the same property. The tool \emph{pState} creates a PRISM formula by automatically taking into account the sates hierarchy.  In a similar way it is possible to specify a reward property by~\emph{"?tran.max"} which is translated into the reward  formula
\[
R\{"tran" \}max=?[F(scopeVariable = state)] 
\]
Properties are specified according to the following grammar:
\[\begin{array}{lll}
Formula ::= "?"  ( \emph{probability} \mid \emph{reward})  ( "."  \mid > \mid < ) ("max" \mid "min" \mid \emph{real} )  ["F < " \; time] 
\\
\emph{probability} ::= "P" \\
\emph{reward} ::= "\$" \emph{identifier}  \\
\emph{time} ::= digit \{digit\} ("d" | "h" | "s" | "ms" | "\mu s") \\
\emph{identifier} ::=  letter \{ letter \mid digit \} \\
\end{array}\]

\noindent Another way to specify the property is to write the formula in the special formula text-box. The properties  
are written in PCTL~\cite{ASB95, KwiatkowskaNormanParker11PRISM4}, a probabilistic extension of temporal logic CTL~\cite{CA82}. The logic for MDP and PTA properties specification is similar, the difference is that PTA property includes clock constraints.  PTA has two model checking engines \emph{digital clocks}~\cite{KNPS06} and \emph{stochastic games}~\cite{PTAKwiatkowska:2009}. In the \emph{digital clock} engine, clock variables are allowed in $P$ operator expressions and temporal logic property types $F$ and $U$ can be used. However, this engine does not support time-bounded reachability properties, as the one used in the section~\ref{sec:CaseStudyHT} example.  

\subsection{Floating Formula Example}
For the system of seven states $\{S0, \ldots, S6\}$ and five transitions $\{t0, \ldots, t4\}$ in Fig.~\ref{fig:FormulaFloatWindow}, we want to find out the minimal probability to reach a particular state. The initial state is $S0$ and probabilistic transition $t0$ moves the system to $S1$ with 30\% probability and to $S2$ with 70\% probability. This is indicated by two alternatives $@0.3$ and $@0.7$ going from the $P$ pseudo-state to the states $S1$ and $S2$. To find out the minimal probability of reaching state $S3$, we need to place the pCharts formula $?P.min$ in the state $S3$. \emph{pState} creates the PRISM formula 
\[
 Pmin=? [ F (root=S3) ]  
\]
which returns 0.03 as the verification result, as can be verified manually by multiplying the probabilities $0.3$ and $0.1$. 

\begin{figure}[!htb] 
\center
\includegraphics[trim = 25mm 201mm 10mm 10mm, clip, width=13.5cm]{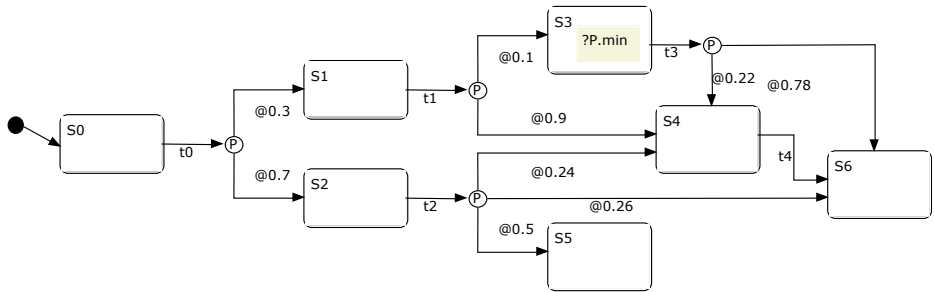}
\caption{Probability of State Reachability}
\label{fig:FormulaFloatWindow}
\end{figure} 

\noindent To determine the probability of entering $S4$, which can be reached on three different paths $S0 \to S1 \to S3 \to S4$, $S0 \to S1 \to S4$, and $S0 \to S2 \to S4$, it is sufficient to move the formula $?P.min$ from state $S3$ to state $S4$, as shown in Figure~\ref{fig:MovingFormula}.  

\begin{figure}[H]
\center\includegraphics[width=0.35\textwidth]{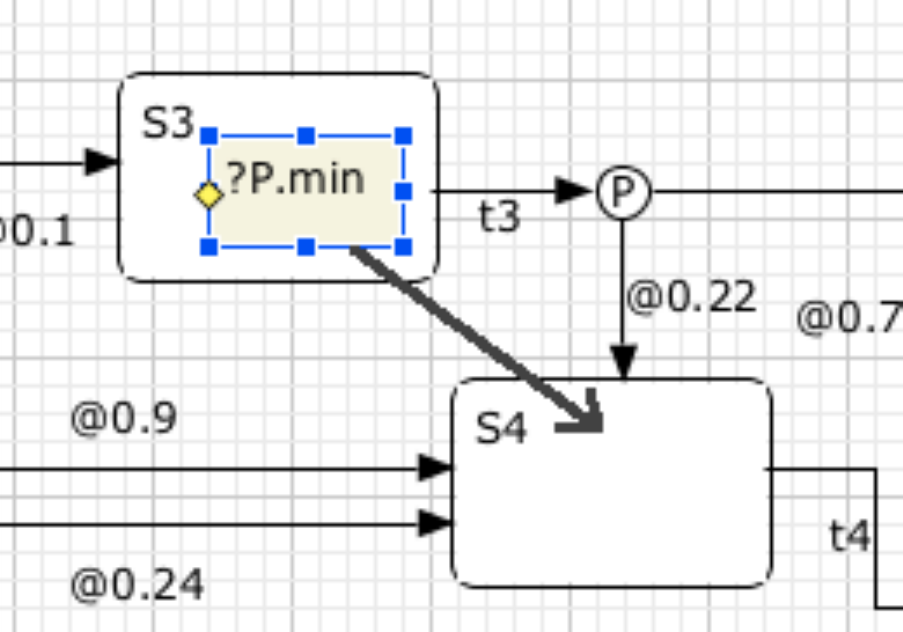}
\caption{Moving Formula From One State to Another}
\label{fig:MovingFormula}
\end{figure}
\noindent Verification of
\[
 Pmin=? [ F (root=S4) ]   
\]
returns that minimal probability to reach state $S4$ is 0.4446. 
Moving formula box to state $S6$, \emph{pState} creates formula 
\[
 Pmin=? [ F (root=S6) ] 
\]
which as result returns 0.6499, that is actually 0.65. The error comes from floating point rounding of the model checker. This would be difficult to calculate manually since there are five different paths to reach this state. 

In this example we calculated $minimal$ probability, but since there are no nondeterministic transitions in the system, the calculation of the $maximal$ probability by $"?P.max"$ would return the same result.

%
\section{Documentation in pCharts}
\label{sec:Comments}
The design can be documented using text boxes in the model itself. The comments are inserted in the generated code. There are three  types of comments \emph{general comments}, \emph{state comments} and \emph{transition comments}. The main reason for including the documentation is to justify the design. Passing the comments to the generated code allows for forward and backward traceability, which would be necessary for the certification of the generated code.

Each comment can be connected to either a state or a transition by dashed comment line. If it is not connected, it is associated to the state surrounding the comment box. General comments are placed outside any states and are technically associated to the root state.

\paragraph{State Documentation}
As the generated code is \emph{event centric}, i.e. states become variables and events become procedures, the comments about a state are inserted in the generated code where the state is declared.  

\paragraph{Transition Documentation}
A transition comment is inserted into generated code of transition event. The comment is connected to the transition by \emph{Comment Connector} figure. Timed transitions do not have an associated event name, but a name is generated for the corresponding procedure and the comment is associated to that procedure in the same way as for untimed events.

\paragraph{ Example}
In Fig.~\ref{fig:Documentation}, a simple transition from state $Off$ to state $On$ on the event $poweron$ is shown. In the grey text-box connected by a dashed line to the state $Off$, the description of the state is given. Another option to describe the state is to place description text-box in the state, as it is shown for $Off$. Code generated for this example is shown in Figure~\ref{fig:GenCode}.
   
\begin{figure}[!htb]
\center
\includegraphics[trim = 25mm 171mm 10mm 10mm, clip, width=9.5cm]{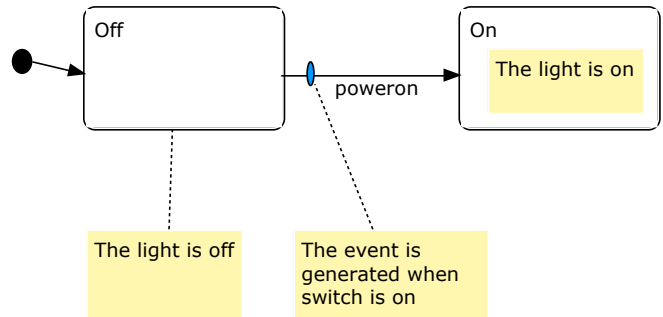}
\caption{State and Transition Documentation}
\label{fig:Documentation}
\end{figure} 

\begin{figure}[!htb]
\begin{lstlisting} [frame=single, mathescape]
/* Variables */
enum root_status {Off, On} root;
// Off - The light is off
// On - The light is on  

int main(void){
    /* Initialization */
    root = Off;

    return 0;
}

// The event is generated when the switch is on
void poweron(void){
    if ((root == Off)) {
        root = On;
    }
}
\end{lstlisting}
\caption{Generated Code for Simple Transition with Comments}
\label{fig:GenCode}
\end{figure}

%
\section{Case Study - RFID Tag}
\label{sec:CaseStudy}
In the pCharts model of Fig.~\ref{fig:PostalSystem} we analyze properties of an RFID tag used in postal systems. The model has two concurrent states, \emph{Tag} and \emph{Environment}. In the \emph{Tag} state the basic operation of the RFID device is represented. Initially, a tag is in \emph{StandBy} and on \emph{wakeUp} goes into \emph{Receive}. This event is broadcasted by \emph{Environment}. The environment is initially in \emph{NoField} and on event \emph{fieldOn} it goes to \emph{SystemField} or to \emph{Interference}. Based on testing we estimate that  approximately 30\% of the time the tag will be excited by a system field and in 70\% of time it will be excited by some unwanted field which may come from some other sources of low frequency like computers, TV, some machinery, etc. On the transition to \emph{SystemField}, the event \emph{wakeUp} is generated. After time \emph{T0} it goes back to \emph{NoField} and increments \emph{counterB}, which counts valid excitations. On the event \emph{T1}, \emph{environment} goes from \emph{Interference} to \emph{NoField} state. We associate the interference cost $interf=1$ to this transition.

On the tag side, in \emph{Receive}, we read the system field ID. If the field ID is recognized, the tag goes into the \emph{Transition}, and if not, it goes back into \emph{StandBy}. We estimate that a tag can recognize the field  ID 80\% of the time. Once it finishes the messages transition, the tag goes back into \emph{Sleep} and increases successful transmissions counter \emph{counterA}. In each state of \emph{Tag}, we specify the costs of being in that state. This is used to evaluate power consumption and for optimization of the system.       

\begin{figure}[tb]
\center\includegraphics[trim = 0mm 150mm 5mm 9mm, clip, width=13.5cm]{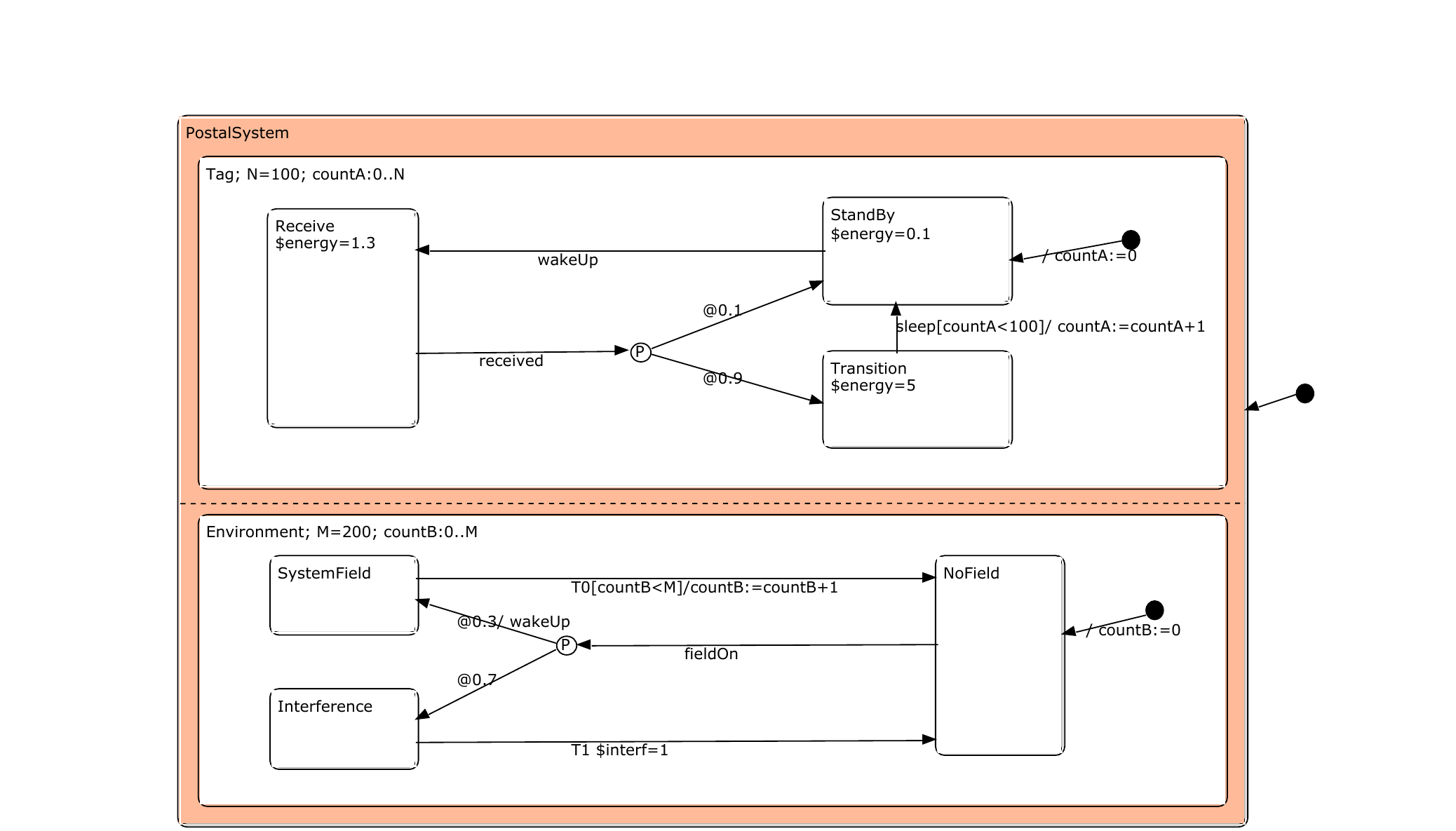}
\caption{Model of RFID Tag Excitation }
\label{fig:PostalSystem}
\end{figure}

\noindent By selecting \emph{View $\to$ Verify} the MDP model of the pCharts is built and passed to the PRISM model checker together with properties. In a separate window the result of property verification is displayed. The model built by PRISM has 136955 states and 318657 transitions. In this example we verify three properties. Those formulas contain conditions of the counters which have to be satisfied and are not specified in the states, but in the formula text box. The the formula
\begin{equation} \label{eq:energ}
?\$energy.min \; (countB=10)
\end{equation}
we can calculate the minimum expected energy on ten \emph{wakeUp} events; the result is: 60.06. The formula
\[
?P.max \; (countA=2) \& (countB<5)
\]
calculates the maximum probability of having two successful transmissions in less than five excitations. The result is 0.99. The formula 
 \[
?\$interf.min \; (countB=10)
 \]
calculates the minimum expected number of interferences in ten good excitations. The result is: 23.33.
 
If we change the probabilities of transitions, or energy consumption we can automatically calculate new values. For instance if the hardware design is improved by the selection of better components such that the  energy in \emph{Receive} is $1.2$, formula~\ref{eq:energ} will result in 59.06; that is by decreasing state consumption for 8.7\% minimum expected energy on ten wakeUp events is reduced by 2.7\%. If costs of new components and labour to alternate receiver is less than saving in energy, that alternation can be done.       

%
\section{Case Study - Hubble Telescope}
\label{sec:CaseStudyHT}
This example is based on the model of failure of the Hubble telescope as presented in~\cite{Oldenkamp07}.
Six gyroscopes are used for navigation. The telescope is designed such that it can fully operate with only three gyroscopes. When less than three gyroscopes are operational,  the telescope goes into sleep mode and waits for repair. As long as at least one gyroscope works, the telescope is operational, otherwise it will crash. The goal of modelling is to find out the probability that the system will operate without failure for a given period of time.

We build a formal probabilistic model of the system as a pCharts model with 13 states. In the model we assume that each gyroscope has an average lifetime of 10 years. Since six gyroscopes are operational, and any one can fail, we can expect that the outgoing rate is~$6 \cdot \frac{1}{10}=0.6$, which means that there is a 60\% of chance that at least one gyroscope will fail in 365 days. That is modelled as a probabilistic timed transition from state \emph{SixG} to state \emph{FiveG}. If five gyroscopes are correct, the probability to have a failure in one year is 0.5, which is modelled as probabilistic transition from state $FiveG$ to state $FourG$. When only two gyroscopes are active, the telescope needs to go into sleep mode. The probability that the telescope will go into sleep mode and the rescue operation starts in three days is 99.8\%. In that case, in approximately 60 days, with probability 0.968\% all failed gyroscopes will be fixed and the system goes into the initial $SixG$ state. If the rescue operation fails, the system goes into $FailOne$ and consecutively into $SleepOne$ state. This means only one gyroscope is functional, and the telescope is in sleep mode. The rescue operation is taken within two months (60 days) with 98.4\% chance of success. While in state $TwoG$, if the system fails to go to $SleepTwo$ state, it will continue to work with two gyroscopes  until one fails in approximately 730 days. Then, it tries to go from $OneG$ into $SleepOne$ and to start the rescue operation. The telescope can end up in the $Crash$ state if it can not go into sleep mode, or if the rescue operation is not successful.
\begin{figure}[tb]
\center\includegraphics[trim = 15mm 150mm 5mm 9mm, clip, width=13.5cm]{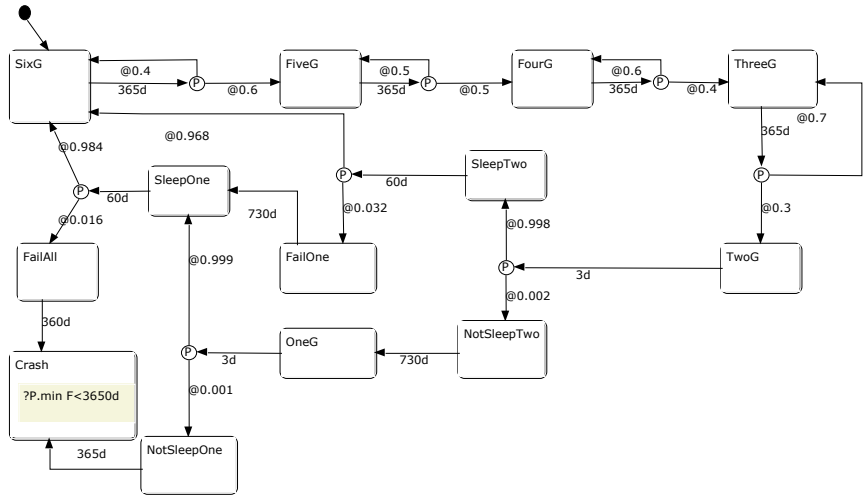}
\caption{pCharts Model of Hubble Space Telescope}
\label{fig:Hubble1}
\end{figure}
From the model, by formula
\[
 Pmin=? [ F <3650 (root=Crash) ]
 \]
 we determine that the crash probability in 10 years (3650 days) is 
 0.01226\%, or the probability that the Hubble telescope will be operational in 10 years is 99.98774\%.

%
\section{Conclusion}
 This paper reports on the formalism of pCharts and its associated tool \emph{pState}. The focus is on properties specification directly on the model in an intuitive way. We believe this technique will make the model checking approach more convenient for developers who are domain experts, but not software experts. The design can be documented and the documentation is passed to the generated code to allow traceability, e.g. for the certification of the generated code. The goal is to have a \emph{seamless} and \emph{automated} approach from modelling and analysis to code generation that can be used to evaluate design alternatives and generate \emph{trustworthy} code. Future work will include formal proof of  correctness of the generated code. 

Our approach is \emph{holistic}, means that qualitative properties, notably structural well-formedness, correctness with respect to invariants, and timing guarantees, can be verified together with quantitative properties, notably resource consumption, reliability, and performance. These properties cannot be analyzed by considering exclusively the computerized part; rather, its environment has to be considered to certain extent. 

\nocite{*}
\bibliographystyle{eptcs}
\bibliography{fdereferences}
\end{document}